# Quantitative evaluation of simultaneous spatial and temporal regularization in liver perfusion studies using low-dose dynamic contrast-enhanced CT

Kenya Murase[a]*, Atsushi Nakamoto[a], Noriyuki Tomiyama[b]

[a]*Department of Future Diagnostic Radiology, Graduate School of Medicine, Osaka University, Suita, Osaka 565-0871, Japan*
[b]*Department of Diagnostic and Interventional Radiology, Graduate School of Medicine, Osaka University, Suita, Osaka 565-0871, Japan*

*Corresponding author:
    Kenya Murase, PhD
    Department of Future Diagnostic Radiology,
    Graduate School of Medicine, Osaka University
    2-2 Yamadaoka, Suita, Osaka 565-0871, Japan
    E-mail: murase@sahs.med.osaka-u.ac.jp

**Abstract:**

**Purpose:** To quantitatively evaluate the performance of different simultaneous spatial and temporal regularizers in liver perfusion studies using low-dose dynamic contrast-enhanced computed tomography (DCE-CT).

**Methods:** A digital liver phantom was used to simulate chronic liver disease (CLD) and hepatocellular carcinoma (HCC) based on clinical data. Low-dose DCE-CT images were reconstructed using regularizers and a primal-dual algorithm. Subsequently, hepatic perfusion parameter (HPP) images were generated using a dual-input single-compartment model and linear least-squares method. In the CLD model, the effect of regularizers on the input functions (IFs) was examined by calculating the areas under the curves (AUCs) of the IFs, and the HPP estimation accuracy was evaluated by calculating the error and coefficient of variation (CV) between the HPP values obtained by the above methods and true values. In the HCC model, the ratios of the mean HPP values inside and outside the tumor were calculated.

**Results:** The AUCs of IFs decreased with increasing regularization parameter (RP) values. Although the AUC of arterial IF did not significantly depend on the regularizers, that of portal IF did. The error and CV were reduced using low-rank and sparse decomposition (LRSD). Total generalized variation (TGV) combined with LRSD (LTGV) was generally superior to the other regularizers in terms of HPP estimation accuracy and range of available RP values in both the CLD and HCC models. However, striped artifacts were more remarkable in the HPP images obtained by the TGV and LTGV than in those obtained by the other regularizers.

**Conclusions:** The results suggest that the LRSD and LTGV are useful for improving the accuracy of HPP estimation using low-dose DCE-CT and for enhancing its practicality. This study will help select a suitable regularizer and/or RP value for low-dose DCE-CT liver perfusion studies.

# 1 INTRODUCTION

The liver is an organ with two blood supplies from the hepatic artery and portal vein. The utility of hepatic perfusion characterization relies on the resolution of each component of its two blood supplies, since contributions from each are altered in many liver diseases.[1] In chronic liver diseases (CLDs), the portal fraction of liver perfusion decreases owing to an increase in intrahepatic vascular resistance due to the progression of hepatic fibrosis.[2] This decrease in portal perfusion is partially compensated by an increase in hepatic arterial perfusion.[2] Consequently, separate measurement of hepatic arterial and portal venous perfusion is important for evaluating the severity of these diseases. Primary and metastatic liver tumors play a major role in mortality and are associated with many cancers. Hepatocellular carcinoma (HCC) is the most common primary hepatic malignancy and one of the most common causes of cancer-related death.[3] Although the underlying mechanism is different from that of CLD, liver tumors also lead to global or regional changes relative to increased arterial fraction and decreased portal fraction of liver perfusion.[3] The noninvasive measurement of these changes is of clinical interest as these changes have been considered an important factor in diagnosing the treatment response of liver tumors.[4]

Dynamic contrast-enhanced computed tomography (DCE-CT) is a promising tool for the noninvasive investigation of hemodynamic changes in tissues and organs.[5] It aids in liver perfusion studies and quantification of hepatic perfusion parameters (HPPs), such as hepatic arterial and portal venous blood flow. Assessing the state of perfusion in the liver using these HPPs is important for determining medical treatment plans and/or predicting prognosis in patients with liver diseases.[1] One of the advantages of using DCE-CT is that the input functions required for the HPP quantification can be obtained noninvasively from high-resolution images.

Radiation exposure during DCE-CT perfusion studies is a serious issue that hinders its widespread clinical use.[6] Lowering the X-ray tube current of CT is one solution for reducing the radiation dose to patients. However, this leads to an increase in the statistical nose in CT images.[7] Thus, various methods, including linear and nonlinear filtering techniques,[8–10] have been proposed to reduce noise in CT images. Recently, denoising methods using artificial intelligence have also been attempted.[11] To facilitate the application of these methods to CT images, supervised and self-supervised approaches using convolutional neural networks have been proposed.[12–14]

With the advent of compressed sensing,[15] sparsity-inducing regularizers, such as the total variation (TV) norm,[16] have been used to enforce piecewise smoothness in images.[17] These regularizers have also been applied to DCE-CT perfusion studies for denoising their images.[18–20] However, the use of the TV often causes staircase artifacts, as the TV minimization forces the flat regions in images to be constant.[16] To reduce the staircase artifacts, the use of higher-order TV (HOTV) has been proposed.[21] Of the regularizers using the HOTV, total generalized variation (TGV)[22] has attracted much attention and has been applied to image denoising and reconstruction.[23–26]

Recently, TGV and its combination with low-rank and sparse decomposition (LRSD)[27] have been applied to cerebral perfusion studies using low-dose DCE-CT.[28] The results suggested its usefulness in improving the accuracy of cerebral perfusion parameter (CPP) estimation and the quality of the CPP images generated from low-dose DCE-CT data.[28] However, to the best of our knowledge, these regularizers have not been applied to liver perfusion studies using low-dose DCE-CT, and their usefulness has yet to be investigated.

This study was aimed at using different simultaneous spatial and temporal regularizers and their combinations with LRSD in low-dose DCE-CT liver perfusion studies and quantitatively investigating their performance in comparison with that without any regularizaers through simulation studies based on clinical data. In this study, we focused on the effects of these regularizers on the accuracy of HPPs estimated from low-dose DCE-CT images using a dual-input single-compartment model in the CLD and HCC models.

## 2 MATERIALS AND METHODS
### 2.1 Image reconstruction
### 2.1.1 Image reconstruction using analytical method

When no regularizers are used, the DCE-CT images are reconstructed using the following equation:

$$\hat{x} = \underset{x}{\mathrm{argmin}}\, \frac{1}{2}\|Ax - b\|_F^2, \tag{1}$$

where $x$ denotes the DCE-CT image matrix comprising $x_i$ ($i = 1,2,\cdots,N_f$), with $x_i$ and $N_f$ being the image matrix at temporal frame $i$ and total number of frames, respectively, $\hat{x}$ is the reconstructed image matrix, $A$ is the system matrix (sampling operator matrix), $b$ is the projection data matrix (sinogram), and $\|\cdot\|_F$ is Frobenius norm. The analytical solution of Equation (1) is expressed as follows:

$$\hat{x} = (A^T A)^{-1} A^T b, \tag{2}$$

where $T$ denotes the transpose of the matrix. The term $(A^T A)^{-1} A^T$ in Equation (2) corresponds to the filtered back-projection (FBP) operator with a Ram–Lak filter.[29] This analytical method is referred to as "FBP."

### 2.1.2 Image reconstruction using TV

When using TV, DCE-CT images are reconstructed as follows:

$$\hat{x} = \underset{x}{\mathrm{argmin}}\, \frac{1}{2}\|Ax - b\|_F^2 + \alpha\|\nabla_3 x\|_1, \tag{3}$$

where $\|\cdot\|_1$ denotes entrywise $\ell_1$ norm, $\nabla_3$ $(=(\nabla_x, \nabla_y, \nabla_t))$ is the three-dimensional gradient operator comprising spatial ($\nabla_x$ and $\nabla_y$) and temporal gradient operators ($\nabla_t$),[23] and the symbol $\alpha$ is a regularization parameter (RP) for balancing the data consistency and sparsity. This method is referred to as "TV."

### 2.1.3 Image reconstruction using combination of LRSD and TV

When DCE-CT images are reconstructed using a combination of LRSD and TV, $x$ is decomposed into low-rank (L) and sparse (S) components, which are obtained by solving the following minimization problem:

$$\min_{L,S}\, \frac{1}{2}\|A(L + S) - b\|_F^2 + \alpha\|\nabla_3 S\|_1 + \beta\|L\|_*, \tag{4}$$

where $L$ and $S$ denote the L and S components of $x$, respectively, $\|L\|_*$ is the nuclear norm (NN) of $L$ (sum of singular values in $L$), and $\beta$ is the RP. In this method, $\hat{x}$ is calculated as the sum of the L and S components obtained by solving Equation (4). This method is referred to as "LTV."

### 2.1.4 Image reconstruction using TGV

When using a TGV of order 2, the DCE-CT images are reconstructed as follows:

$$\hat{x} = \underset{x}{\mathrm{argmin}}\, \frac{1}{2}\|Ax - b\|_F^2 + TGV_\alpha^2(x), \tag{5}$$

where $TGV_\alpha^2$ denotes the TGV of order 2 and is defined as follows:[24]

$$TGV_\alpha^2(x) = \min_v \alpha_1\|\nabla_3 x - v\|_1 + \alpha_0\|\mathcal{E}(v)\|_1, \tag{6}$$

where $\alpha_0$ and $\alpha_1$ denote the RPs; $\mathcal{E}(v)$ denotes the symmetric gradient of $v$ and is expressed as $\mathcal{E}(v) = (\nabla_3 v + \nabla_3 v^T)/2$. This method is referred to as "TGV."

### 2.1.5 Image reconstruction using combination of LRSD and TGV

As in LTV, when using a combination of LRSD and TGV, $x$ is decomposed into L and S components, and $\hat{x}$ is calculated as the sum of the L and S components obtained by solving the following minimization problem:[23]

$$\min_{L,S} \frac{1}{2}\|A(L+S)-b\|_F^2 + TGV_\alpha^2(S) + \beta\|L\|_*. \qquad (7)$$

This method is referred to as "LTGV."

### 2.1.6 Image reconstruction using combination of first- and second-order TVs

For comparison, we studied a linear combination of first- and second-order TVs. The reconstruction of the DCE-CT images using this approach is formulated as follows:

$$\hat{x} = \underset{x}{\mathrm{argmin}}\, \frac{1}{2}\|Ax-b\|_F^2 + \alpha_1\|\nabla_3 x\|_1 + \alpha_2\|\mathcal{E}(\nabla_3 x)\|_1, \qquad (8)$$

where $\alpha_1$ and $\alpha_2$ denote the RPs. This method is referred to as "FSOTV."

### 2.1.7 Image reconstruction using combination of LRSD and FSOTV

As in LTV and LTGV, when using a combination of LRSD and FSOTV, $x$ is decomposed into L and S components, and $\hat{x}$ is calculated as the sum of the L and S components obtained by solving the following minimization problem:[23]

$$\min_{L,S} \frac{1}{2}\|A(L+S)-b\|_F^2 + \alpha_1\|\nabla_3 S\|_1 + \alpha_2\|\mathcal{E}(\nabla_3 S)\|_1 + \beta\|L\|_*. \qquad (9)$$

This method is referred to as "LFSOTV."

### 2.1.8 Implementation of image reconstruction

To obtain $\hat{x}$ by solving Equations (3)–(9), the primal-dual (PD) algorithm[30,31] was used. The details of the PD algorithm are described by Murase et al.[28] The iterative procedure of the PD algorithm was repeated until $\|\hat{x}^{n+1}-\hat{x}^n\|_F / \|\hat{x}^n\|_F$ became less than $10^{-6}$, or the number of iterations reached 500, where $\hat{x}^n$ denotes the reconstructed image matrix at iteration $n$. When TV, TGV, and FSOTV were used, the initial estimate ($\hat{x}^1$) was replaced by the image matrix obtained by FBP. When using LTV, LTGV, and LFSOTV, the L component of $\hat{x}^1$ was replaced by the image matrix obtained by FBP, whereas the S component was set to a zero matrix.

The RP ($\alpha$ in TV and LTV and $\alpha_1$ in TGV, LTGV, FSOTV, and LFSOTV) values were varied between 0.0001 and 0.01, whereas $\alpha_0$ in TGV and LTGV was set to $\alpha_1 \times 2$, as recommended by Knoll et al.,[25] $\alpha_2$ in FSOTV and LFSOTV was set to $\alpha_1 \times 0.1$, and $\beta$ was set to 2.0. Both the first- and second-order step sizes in the PD algorithm[30,31] were set to 0.25. These values were determined via trial and error using hand tuning with reference to previous studies[23,28,32] and by considering the convergence behavior and speed.

## 2.2 HPP estimation using dual-input single-compartment model

When using a dual-input single-comparment model and assuming that the contribution of the contrast agent (CA) in the intrahepatic artery and portal vein is negligible, the time-dependent concentration of CA in the liver region of interest (ROI) ($C_L(t)$) is expressed as[2]

$$C_L(t) = K_{1a} \cdot \int_0^t C_a(\tau - \tau_a) \cdot e^{-k_2(t-\tau)} d\tau + K_{1p} \cdot \int_0^t C_p(\tau - \tau_a) \cdot e^{-k_2(t-\tau)} d\tau. \tag{10}$$

In Equation (10), $C_a(t)$ and $C_p(t)$ denote the time-dependent concentrations of the CA in the hepatic artery and portal vein, that is, hepatic arterial and portal venous input functions, respectively, and $\tau_a$ and $\tau_p$ denote the transit times of the CA from the hepatic artery and portal vein to the liver ROI, respectively. When deriving Equation (10), $C_a(t)$ and $C_p(t)$ were assumed to be zero at $t \leq \tau_a$ and $t \leq \tau_p$, respectively. $K_{1a}$, $K_{1p}$, and $k_2$ are the rate constants for the exchanges of CA from the hepatic artery to the liver, portal vein to the liver, and liver to blood, respectively. Notably, $K_{1a}$ and $K_{1p}$ are equal to $E \times F_a$ and $E \times F_p$, respectively, where $E$, $F_a$, and $F_p$ denote the hepatic extraction fraction, hepatic arterial blood flow, and portal venous blood flow, respectively. The reciprocal of $k_2$ corresponds to the mean trnasit time (MTT) of CA in the liver.[2]

When uing the relation $C_a(t - \tau_a) \otimes e^{-k_2 t} = \int_0^t C_a(\tau - \tau_a) d\tau - k_2 \cdot \int_0^t C_a(\tau - \tau_a) \otimes e^{-k_2 \tau} d\tau$ ($\otimes$: convolution integral), Equation (10) is reduced to

$$C_L(t) = K_{1a} \cdot \int_0^t C_a(\tau - \tau_a) d\tau + K_{1p} \cdot \int_0^t C_p(\tau - \tau_p) d\tau - k_2 \cdot \int_0^t C_L(\tau) d\tau. \tag{11}$$

When Equation (11) is expressed in discrete form, we obtain the following matrix equation:

$$\begin{pmatrix} C_L(t_1) \\ C_L(t_2) \\ \vdots \\ C_L(t_{N_f}) \end{pmatrix} = \tag{12}$$

$$\begin{pmatrix} \int_0^{t_1} C_a(\tau - \tau_a)d\tau & \int_0^{t_1} C_p(\tau - \tau_p)d\tau & -\int_0^{t_1} C_L(\tau)d\tau \\ \int_0^{t_2} C_a(\tau - \tau_a)d\tau & \int_0^{t_2} C_p(\tau - \tau_p)d\tau & -\int_0^{t_2} C_L(\tau)d\tau \\ \vdots & \vdots & \vdots \\ \int_0^{t_{N_f}} C_a(\tau - \tau_a)d\tau & \int_0^{t_{N_f}} C_p(\tau - \tau_p)d\tau & -\int_0^{t_{N_f}} C_L(\tau)d\tau \end{pmatrix} \begin{pmatrix} K_{1a} \\ K_{1p} \\ k_2 \end{pmatrix}.$$

where $t_i$ ($i = 1, 2, \cdots, N_f$) denotes the middle frame time of the *i*th frame. When $C_L(t_i)$, $C_a(t_i - \tau_a)$, and $C_p(t_i - \tau_p)$ ($i = 1, 2, \cdots, N_f$) are known, the HPPs ($K_{1a}$, $K_{1p}$, and $k_2$) can be easily estimated from Equation (12) using the linear least-squares (LLSQ) method.[33] The LLSQ method requires no initial estimates and its computational cost is much less than that of the nonlinear least-squares method.[33]

In X-ray CT, the concentration of CA is approximated to be proportional to an increase in the CT number, that is, contrast enhancement (CE),[34] and $C_L(t)$, $C_a(t)$, and $C_p(t)$ in Equation (10) can be replaced by CE at time *t* in the liver, hepatic artery, and portal vein.

## 2.3 Simulation study
### 2.3.1 Generation of DCE-CT and HPP images

In this study, a digital liver phantom was constructed based on DCE-CT images acquired from a 71-year-old female patient according to the following protocol:[35] After obtaining informed consent, DCE-CT images with a matrix size of 512×512 were acquired using multidetector row CT (Aquilion 64; Canon Medical Systems Co., Tochigi, Japan) with a tube voltage and current of 120 kVp and 120 mA, respectively. A total of 21 scans comprising four contiguous 8-mm-thick slices at the porta hepatica level were acquired with a gantry rotation speed of 1 s per rotation, 7 s after intravenously injecting 40-mL iodinated CA (Iopamiron 370; Bayer Yakuhin, Ltd., Osaka, Japan) at 5 mL/s through a 20-gauge indwelling needle in the antecubital vein. Fifteen scans with a duration of 2 s were acquired for 30 s, followed by six scans with a duration of 7 s. Two intervals with a duration of 9.6 s were taken between the 15th and 16th and the 18th and 19th scans. The patient was instructed to hold her breath during scanning. The obtained DCE-CT images were then interpolated to generate 44 images per slice for a duration of 2 s using the linear interpolation technique, and the matrix size was reduced to 256×256.

Figure 1 illustrates the ROIs of the liver (yellow), spleen (green), aorta (red), and portal vein (pink) in the CT image. The black solid circle shows the tumor boundary (a circle with a diameter of 25 pixels) considered in the HCC model. The red, blue, and green solid lines in Figure 2 show the CE curves in the ROIs of the aorta, portal vein, and spleen, respectively (Figure 1). A digital liver phantom was constructed by replacing the CE curves in the liver ROI with $C_L(t)$ calculated from Equation (10) with varying $K_{1a}$, $K_{1p}$, and $k_2$ values. The black solid line in Figure 2 shows an example of $C_L(t)$ with $K_{1a}$ = 20 mL/100mL/min, $K_{1p}$ = 80 mL/100mL/min, $k_2$ = 4 min$^{-1}$, and $\tau_a = \tau_p = 0$ s. The pixel intensity of the digital liver phantom was in Hounsfield unit (HU).

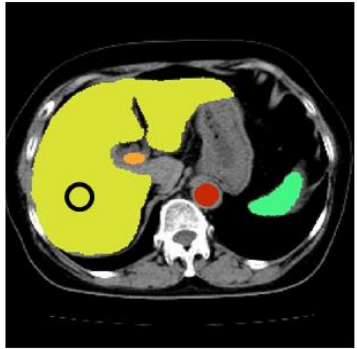

**Fig. 1.** Illustration of the regions of interest (ROIs) drawn on the CT image acquired from a patient. Yellow, green, red, and pink regions illustrate the ROIs of the liver, spleen, aorta, and portal vein, respectively. Black solid circle shows the boundary of the tumor ROI in the hepatocellular carcinoma (HCC) model.

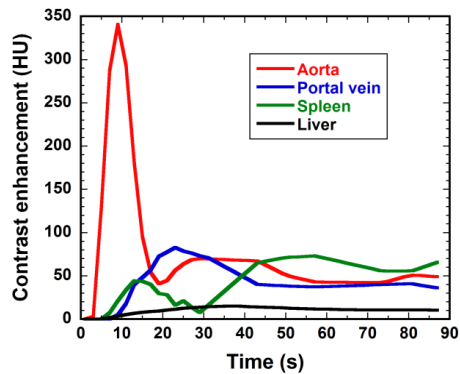

**Fig. 2.** Contrast enhancement (CE) curves in the aorta (red), portal vein (blue), and spleen ROIs (green) obtained from a patient (Figure 1). Black solid curve shows an example of the CE curve in the liver ROI, which was calculated from Equation (10) assuming that $K_{1a}$ = 20 mL/100mL/min, $K_{1p}$ = 80 mL/100mL/min, $k_2$ = 4 min$^{-1}$, and $\tau_a = \tau_p = 0$ s. The CE curves in the aorta and portal vein ROIs were used as the true input functions. HU: Hounsfield unit.

The projection data matrix (**b** in Equation (1)) for DCE-CT images were generated from the linear attenuation coefficient ($\mu$) maps transformed from the digital liver phantom using a simple monoenergetic forward model assuming the effective X-ray energy for a tube voltage of 120 kVp of 56 keV[36] and the $\mu$ values for air (−1000 HU) and water (0 HU) of 0 and 0.212 cm$^{-1}$, respectively. The $\mu$ value for water was obtained from the photon cross section data provided by XCOM (https://physics.nist.gov/PhysRefData/Xcom/). In this study, we considered a fan-beam projection geometry[37] with 1200 projections equally spaced over a 360° angular range. The X-ray detector was arranged in an arc shape with 975 sensors. The source-to-isocenter and source-to-detector distances were 539 and 947 mm, respectively. The field of view for image reconstruction was 309 mm×309 mm (1 pixel = 1.21 mm×1.21 mm). The incident X-ray photons were assumed to pass through an object according to the Beer–Lambert law, that is, the number of surviving photons (*I*) was calculated from $I = I_0 e^{-P}$, where $I_0$ and *P* denote the number of incident photons and line integral of the $\mu$ values along the projection ray passing through the object, respectively. Unless otherwise stated, $I_0$ was assumed to be 5×10$^5$ for each ray. Poisson noise was added to the projection data matrix to simulate statistical noise. In addition, readout Gaussian noise with a variance of 10 photons was added. When *I* reached zero or became negative, it was clamped to unity to avoid logarithmic zero or negative values. The system matrix (***A*** in Equation (1)) was calculated using Siddon's algorithm.[38]

The DCE-CT images were reconstructed from the projection data matrix obtained using Equations (1)–(9). CE images were generated by calculating the average of each pixel intensity of the DCE-CT images between the first and second frames (baseline) and subtracting them from each pixel intensity. HPP images were generated from these CE images by solving Equation (12) pixel by pixel using the LLSQ method[33]. In this study, $C_a(t)$ was replaced by the CE curve in the aorta, and $\tau_a$ and $\tau_p$ were assumed to be zero for simplicity.

**2.3.2 Area under the curve of input function**

To investigate the effect of the input functions on the HPP estimation accuracy, we studied three types of input functions: 1) the true input functions (Figure 2), those extracted from 2) the DCE-CT images reconstructed using different regularizers and 3) images obtained by FBP. In this study, these input functions were referred to as "true," "individual," and "FBP" input functions, respectively. To evaluate the effect of these input functions quantitatively, the area under the curve (AUC) was calculated by integrating the input function from 0 to the last time point using the trapezoidal rule. Furthermore, the normalized AUCs were calculated by dividing the AUCs by those of the true input functions. Unless otherwise specified, FBP input functions were used.

### 2.3.3 CLD model

To investigate the HPP estimation accuracy, we performed simulations using the CLD model with three different severities: 1) non-cirrhotic CLD, 2) Child A cirrhosis, and 3) Child C cirrhosis, where Child A and Child C refer to the Child–Pugh classification.[2] According to Miyazaki et al.,[39] the HPP values summarized in Table 1a were used as the ground truth in the CLD model.

### 2.3.4 HCC model

In addition to the CLD model, we used the following HCC model: Chen et al.[40] reported that the ratio of hepatic arterial blood flow to total blood flow in HCCs was 0.58 on average and that the total blood flow in HCCs was approximately 1.57 times lower than that in the surrounding liver parenchyma. Based on these data,[40] the $K_{1a}$ and $K_{1p}$ values in HCCs were calculated from those in the surrounding liver parenchyma (denoted by $K_{1a}^s$ and $K_{1p}^s$, respectively) as $(K_{1a}^s+K_{1p}^s)/1.57\times0.58$ and $(K_{1a}^s+K_{1p}^s)/1.57\times0.42$, respectively. Sahani et al.[41] reported that MTT in HCCs was 1.67 times lower than that in background livers with cirrhosis. Since $k_2$ is the reciprocal of MTT, the $k_2$ value in HCCs was calculated as $k_2^s \times 1.67$, where $k_2^s$ denotes the $k_2$ value in the surrounding liver parenchyma. In this study, the surrounding liver parenchyma was assumed to be a non-tumor liver with Child A cirrhosis (Table 1a). Table 1b summarizes the HPP values inside and outside the tumor that were used as the ground truth for the HCC model.

To quantitatively investigate the HPP estimation accuracy of the HCC model, the ratios of the mean HPP values inside and outside the tumor (tumor I/O ratios) were calculated. The true tumor I/O ratios are listed in Table 1b.

TABLE 1 Summary of hepatic perfusion parameter ($K_{1a}$, $K_{1p}$, and $k_2$) values used as ground truth in chronic liver disease (CLD) and hepatocellular carcinoma (HCC) models.

(a) CLD model

| CLD | $K_{1a}$ (mL/100mL/min) | $K_{1p}$ (mL/100mL/min) | $k_2$ (min$^{-1}$) |
|---|---|---|---|
| Non cirrhotic | 20.0 | 80.0 | 4.0 |
| Child A cirrhosis | 20.0 | 50.0 | 3.0 |
| Child C cirrhosis | 40.0 | 10.0 | 2.0 |

(b) HCC model

| HCC | $K_{1a}$ (mL/100mL/min) | $K_{1p}$ (mL/100mL/min) | $k_2$ (min$^{-1}$) |
|---|---|---|---|
| Inside tumor | 25.9 | 18.7 | 5.0 |
| Outside tumor | 20.0 | 50.0 | 3.0 |
| Tumor I/O ratio[a] | 1.30 | 0.37 | 1.67 |

[a]Tumor I/O ratio is the ratio of the HPP values inside and outside the tumor.

### 2.3.5 Error analysis of HPP

The errors were analyzed to quantitatively evaluate the accuracy of the HPP values obtained from the DCE-CT images reconstructed using different methods in the CLD model. The *Error* of HPP $\theta$ ($\theta$: $K_{1a}$, $K_{1p}$, and $k_2$) was calculated as

$$Error = \frac{1}{N_L}\sum_{i=1}^{N_L} \frac{|\theta_i - \theta_{gt}|}{\theta_{gt}} \times 100 \ (\%), \tag{13}$$

where $\theta_i$, $\theta_{gt}$, and $N_L$ denote $\theta$ at pixel $i$, the ground truth of $\theta$ (Table 1a), and the total number of pixels in the liver ROI, respectively, and $|\cdot|$ is the absolute value. In addition, the coefficient of variation (*CV*) was calculated as

$$CV = \frac{SD(\theta)}{\theta_{gt}} \times 100 \ (\%), \tag{14}$$

where $SD(\theta)$ denotes the standard deviation (SD) of $\theta$ in the liver ROI.

### 2.3.6 Statistical analysis

To quantitatively evaluate the effect of input functions, the HPP values obtained using different input functions were expressed as the mean ± SD in the liver ROI. Differences in these values among groups with different input functions were analyzed using one-way analysis of variance. Statistical significance was determined using Tukey's multiple comparison test, and $p < 0.05$ was considered statistically significant.

## 3 RESULTS

The left panel in Figure 3 shows the normalized AUCs of $C_a(t)$ extracted from the DCE-CT images reconstructed using FBP (black solid line), TV (red solid line), LTV (red dotted line), TGV (blue solid line), LTGV (blue dotted line), FSOTV (green solid line), and LFSOTV (green dotted line) as functions of RP ($\alpha$ in TV and LTV and $\alpha_1$ in TGV, LTGV, FSOTV, and LFSOTV) values. The right panel shows those of $C_p(t)$. The AUCs for both $C_a(t)$ and $C_p(t)$ decreased with increasing RP values. Those of $C_a(t)$ did not significantly depend on the regularizers unlike those of $C_p(t)$. The AUC of $C_p(t)$ for LTV decreased the most with increasing RP value, followed by those for LFSOTV, TV, FSOTV, LTGV, and TGV in decreasing order. When using LRSD, the AUCs of $C_p(t)$ were smaller than those without LRSD.

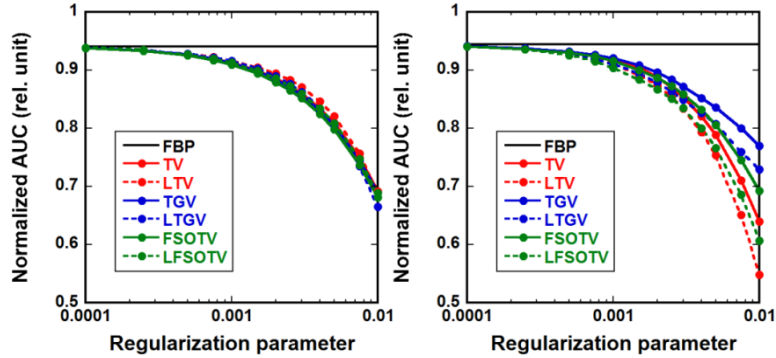

**Fig. 3.** Areas under the curves (AUCs) of the input functions extracted from the DCE-CT images reconstructed using different regularizers (individual input functions) as functions of regularization parameter (RP) ($\alpha$ in TV and LTV and $\alpha_1$ in TGV, LTGV, FSOTV, and LFSOTV) values in non-cirrhotic CLD (Table 1a). The AUCs are normalized by those of the true input functions (Figure 2). Black solid lines show those for the input functions extracted from the DCE-CT images obtained by FBP (FBP input functions). The left and right panels show cases for arterial ($C_a(t)$) and portal venous input functions ($C_p(t)$), respectively. The number of incident photons ($I_0$) was assumed to be $5\times10^5$.

The first, second, and third rows in Figure 4a show the $K_{1a}$ images obtained using the true, individual, and FBP input functions, respectively. They were generated from DCE-CT images reconstructed using TV, LTV, TGV, LTGV, FSOTV, and LFSOTV with an RP value of 0.002 (left to right columns) in non-cirrhotic CLD (Table 1a). Figures 4b and 4c show $K_{1p}$ and $k_2$ images, respectively. As shown in Figure 4, no significant differences were observed among the groups with different input functions in the $K_{1a}$ and $k_2$ images. By contrast, when using the individual input functions, the $K_{1p}$ images exhibited higher pixel values than those for the true and FBP input functions. This was remarkable when using the LRSD.

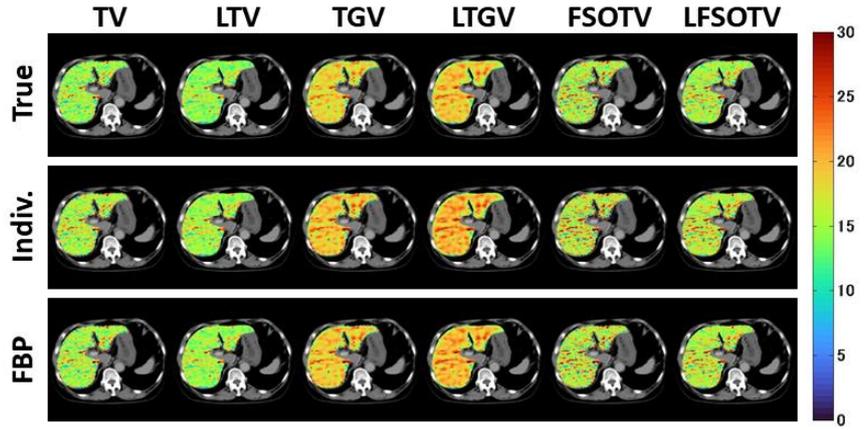

(a)

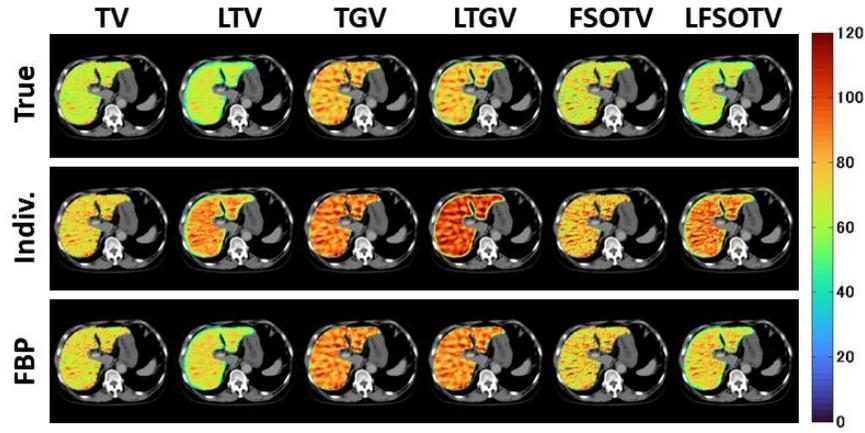

(b)

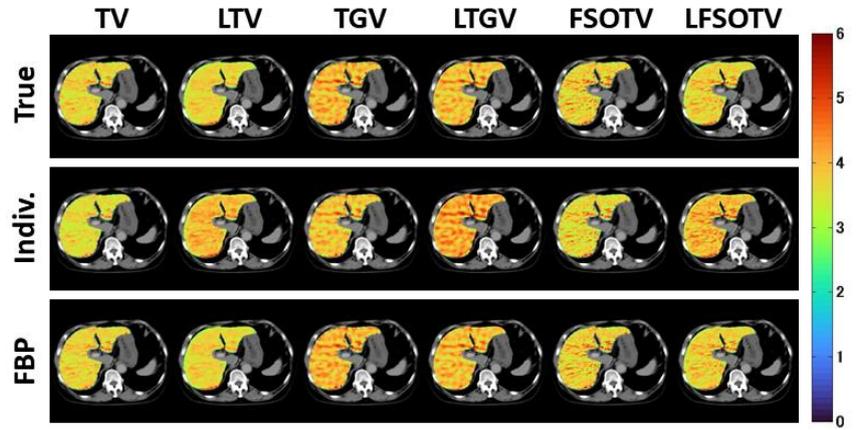

(c)

**Fig. 4.** (a) $K_{1a}$, (b) $K_{1p}$, and (c) $k_2$ images generated from the DCE-CT images reconstructed using TV, LTV, TGV, LTGV, FSOTV, and LFSOTV (left to right columns) with an RP value of 0.002 in non-cirrhotic CLD (Table 1a). The first, second, and third rows show the cases of using the true, individual, and FBP input functions, respectively. The $K_{1a}$, $K_{1p}$, and $k_2$ images are represented in mL/100mL/min, mL/100mL/min, and min$^{-1}$, respectively, and are superimposed on CT images. $I_0$ was assumed to be $5\times10^5$.

Figure 5 shows the $K_{1a}$, $K_{1p}$, and $k_2$ values (left to right panels) obtained from the DCE-CT images reconstructed using TV (red), LTV (red hatching), TGV (blue), LTGV (blue hatching), FSOTV (green), and LFSOTV (green hatching) in non-cirrhotic CLD (Table 1a). The left, middle, and right groups in each panel show the cases of the true, individual, and FBP input functions, respectively. The bars and error bars represent the mean and SD of the liver ROI, respectively. The $K_{1a}$ and $K_{1p}$ values obtained using the individual input functions were significantly higher than those obtained using the true and FBP input functions, and those obtained using the FBP input functions were significantly higher than those for the true input functions in all regularizers (Figures 5a and 5b). When using TV, TGV, and FSOTV, the $k_2$ values obtained using the individual input functions were significantly lower than those obtained using the true and FBP input functions, whereas no significant differences were observed between the groups for the true and FBP input functions. When using LTV, LTGV, and LFSOTV, the $k_2$ values obtained using the individual input functions were significantly higher than those obtained using the true and FBP input functions, and those obtained using the FBP input functions were significantly higher than those obtained using the true input functions (Figure 5c).

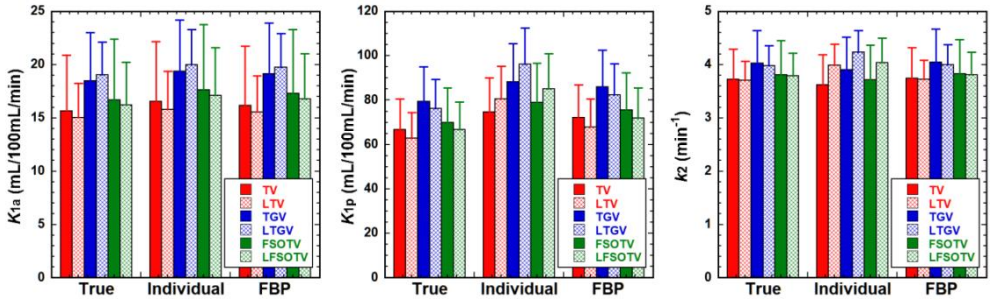

**Fig. 5.** $K_{1a}$, $K_{1p}$, and $k_2$ values (left to right panels) obtained from the DCE-CT images reconstructed using TV (red), LTV (red hatching), TGV (blue), LTGV (blue hatching), FSOTV (green), and LFSOTV (green hatching) in non-cirrhotic CLD (Table 1a). The left, middle, and right groups in each panel show cases when using the true, individual, and FBP input functions, respectively. The bar and error bar represent the mean and SD in the liver ROI, respectively. $I_0$ was assumed to be $5 \times 10^5$.

Figure 6a shows the $K_{1a}$, $K_{1p}$, and $k_2$ images (left to right columns) generated from the DCE-CT images reconstructed using FBP for Child A cirrhosis (Table 1a). The upper and lower rows show cases without and with noise, respectively. Figures 6b, 6c, and 6d show the $K_{1a}$, $K_{1p}$, and $k_2$ images, respectively, obtained using TV, LTV, TGV, LTGV, FSOTV, and LFSOTV (top to bottom rows) with RP values of 0.001, 0.002, 0.003, 0.004, and 0.005 (left to right columns). The $K_{1a}$ images obtained by TGV and LTGV were almost constant at $\alpha_1 \geq$ 0.002, whereas those for the other regularizers decreased with increasing RP values (Figure 6b). The $K_{1p}$ and $k_2$ images did not change significantly at RP values $\geq$ 0.002 (Figures 6c and 6d).

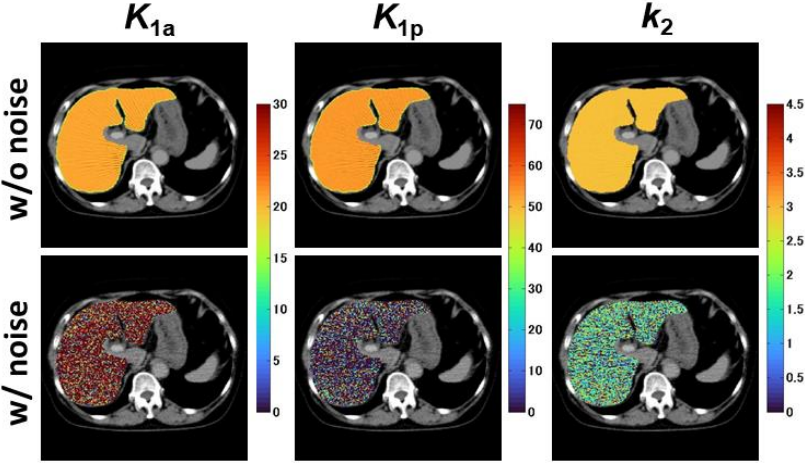

(a)

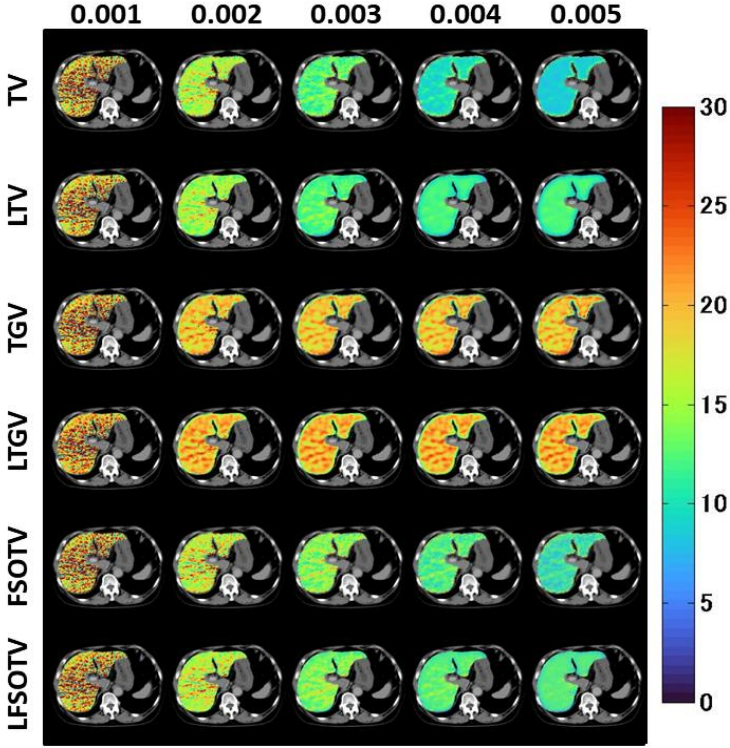

(b)

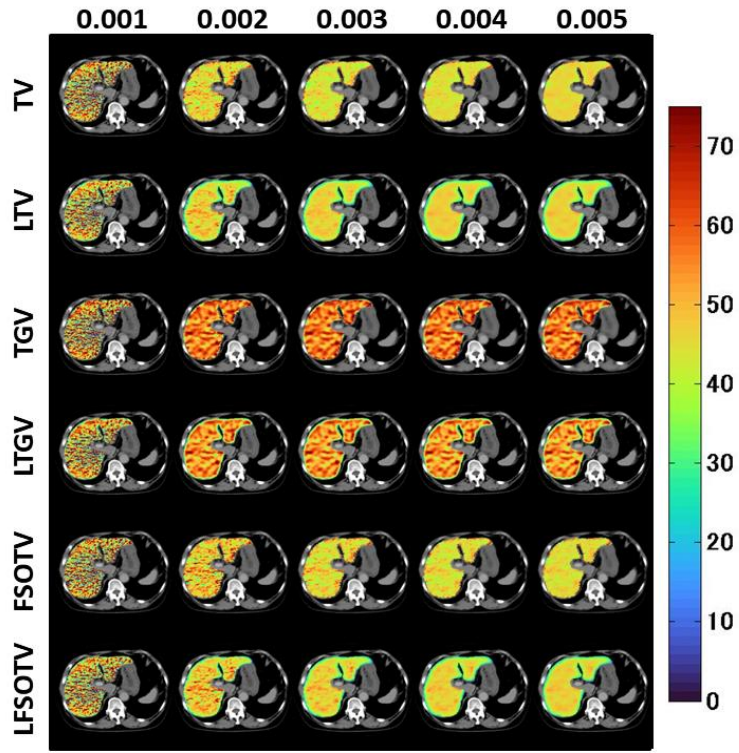

(c)

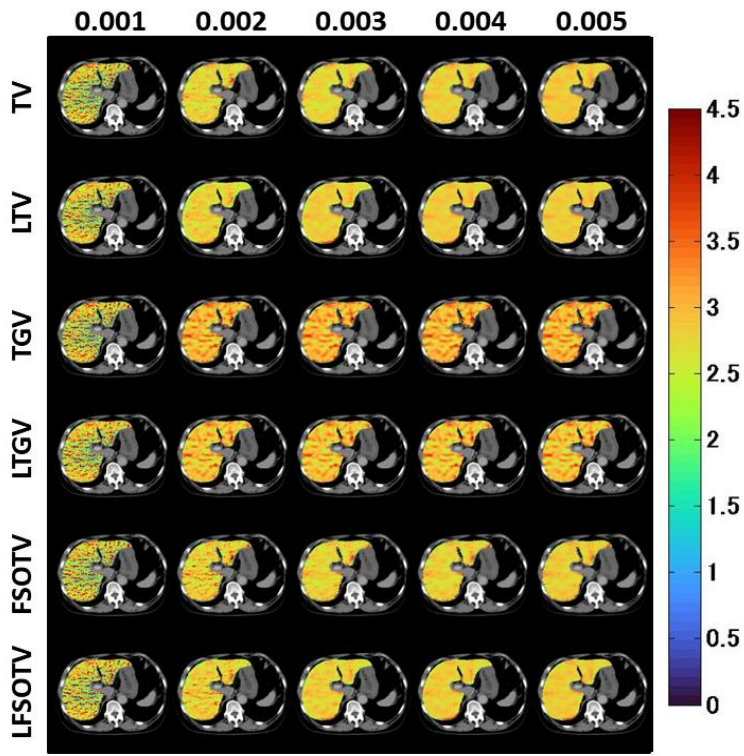

(d)

**Fig. 6.** (a) $K_{1a}$, $K_{1p}$, and $k_2$ images (left to right columns) generated from the DCE-CT images reconstructed using FBP in Child A cirrhosis (Table 1a). The upper and lower rows show cases without and with noise, respectively. (b) $K_{1a}$, (c) $K_{1p}$, and (d) $k_2$ images obtained by TV,

LTV, TGV, LTGV, FSOTV, and LFSOTV (top to bottom rows) with RP values of 0.001, 0.002, 0.003, 0.004, and 0.005 (left to right columns). The $K_{1a}$, $K_{1p}$, and $k_2$ images are represented in mL/100mL/min, mL/100mL/min, and min$^{-1}$, respectively, and are superimposed on CT images. $I_0$ was assumed to be $5\times10^5$, and FBP input functions were used.

The error analysis results are shown in Figure 7. Figure 7a shows the *Error* (Equation (13)) (upper row) and *CV* (Equation (14)) (lower row) of the estimated $K_{1a}$ (red closed circles), $K_{1p}$ (blue closed circles), and $k_2$ values (green closed circles) as functions of RP values in the non-cirrhotic CLD group. Figures 7b and 7c show the cases of Child A and Child C cirrhosis (Table 1a), respectively. The left, middle, and right panels show the cases of TV (solid line) and LTV (dotted line), TGV (solid line) and LTGV (dotted line), and FSOTV (solid line) and LFSOTV (dotted line). The solid lines show the FBP cases. In the non-cirrhotic CLD (Figure 7a), the *Error* and *CV* of $K_{1a}$ were the largest, followed by those of $K_{1p}$ and $k_2$ in decreasing order. When using TV (left column), the *Errors* of $K_{1a}$ and $K_{1p}$ were minimized at $\alpha$ = 0.0015–0.002, following which they increased with increasing $\alpha$. The *Error* of $k_2$ decreased with increasing $\alpha$ until 0.002, following which it plateaued. When using LTV, the *Error* of $K_{1a}$ was smaller than that of TV at $\alpha > 0.003$. When using TGV (middle column), the *Errors* of all HPPs decreased with increasing $\alpha_1$ until 0.002, following which they plateaued. When using the LTGV, the *Errors* were slightly smaller than those for the TGV. When using FSOTV and LFSOTV (right column), the dependencies of the *Error* on RP values were similar to those for TV and LTV (left column), whereas those of the *CV* were similar to those for TGV and LTGV (middle column).

In the case of Child A cirrhosis (Figure 7b), the *Errors* of $K_{1a}$ and $K_{1p}$ were close at $\alpha$ or $\alpha_1$ < 0.0015–0.002, and that of $k_2$ was the smallest. The dependency of *CV* on $\alpha$ or $\alpha_1$ was similar to that in the non-cirrhotic CLD (Figure 7a). When using TV (left column), the *Error* of $K_{1a}$ was minimized at $\alpha$ = 0.0015–0.002, following which it increased with increasing $\alpha$. This increase was slightly suppressed using LTV at $\alpha > 0.003$. The *Errors* of $K_{1p}$ and $k_2$ decreased gradually with increasing $\alpha$. As in the non-cirrhotic CLD (Figure 7a), when using TGV and LTGV (middle column), the *Errors* of all HPPs decreased with increasing $\alpha_1$ until 0.002, following which they plateaued. When using FSOTV and LFSOTV (right column), the dependencies of the *Error* on RP values were similar to those for TV and LTV (left column), whereas those of the *CV* were similar to those for TGV and LTGV (middle column).

In the case of Child C cirrhosis (Figure 7c), the *Error* and *CV* of $K_{1p}$ were much larger than those of $K_{1a}$ and $k_2$. When using TV (left column), the *Errors* of all HPPs were minimized at $\alpha$ = 0.0015–0.002, following which the *Error* of $K_{1p}$ increased steeply. This increase was suppressed using LTV at $\alpha \geq 0.005$. The *CV* was also slightly reduced using LTV. When using TGV (middle column), the *Error* of $K_{1p}$ was minimized at $\alpha_1$ = 0.0015, following which it increased gradually with increasing $\alpha_1$. When using LTGV, the *Error* of $K_{1p}$ plateaued at $\alpha_1 \geq 0.0015$, and the *CV* was smaller than that for TGV. When using FSOTV and LFSOTV (right column), the results were similar to those for TV and LTV (left column).

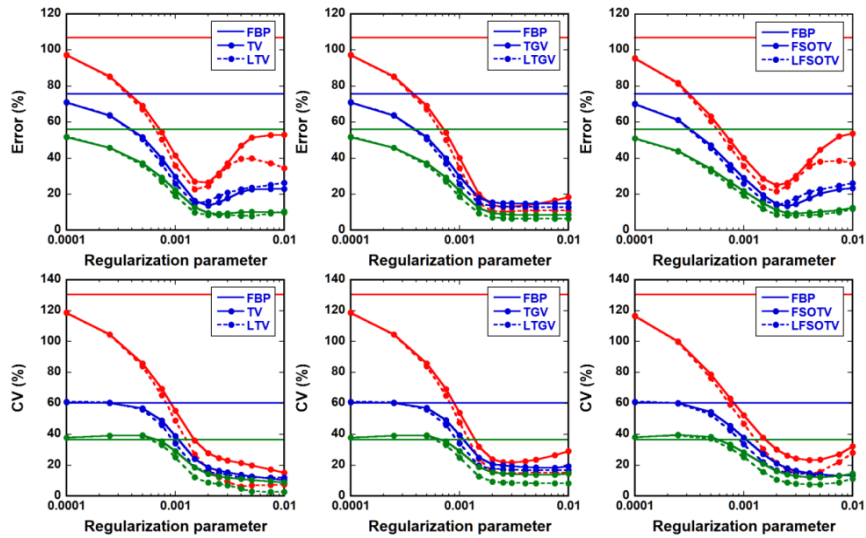

(a)

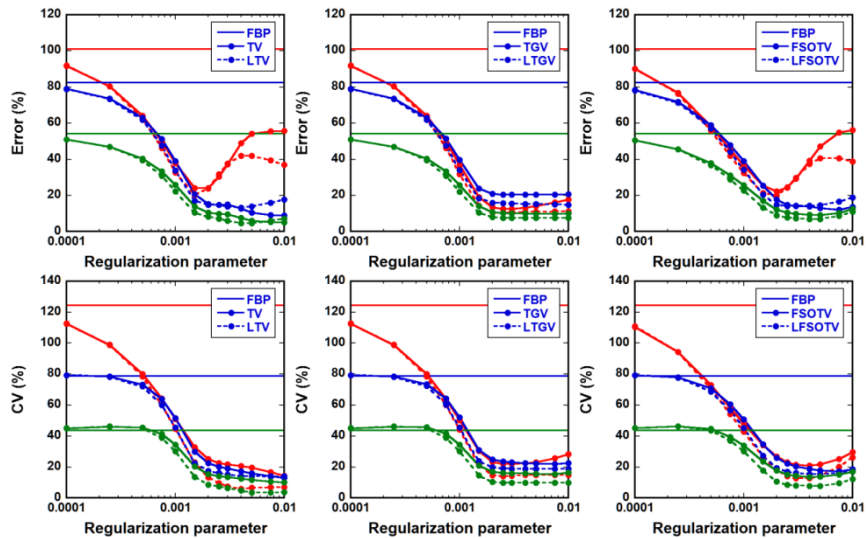

(b)

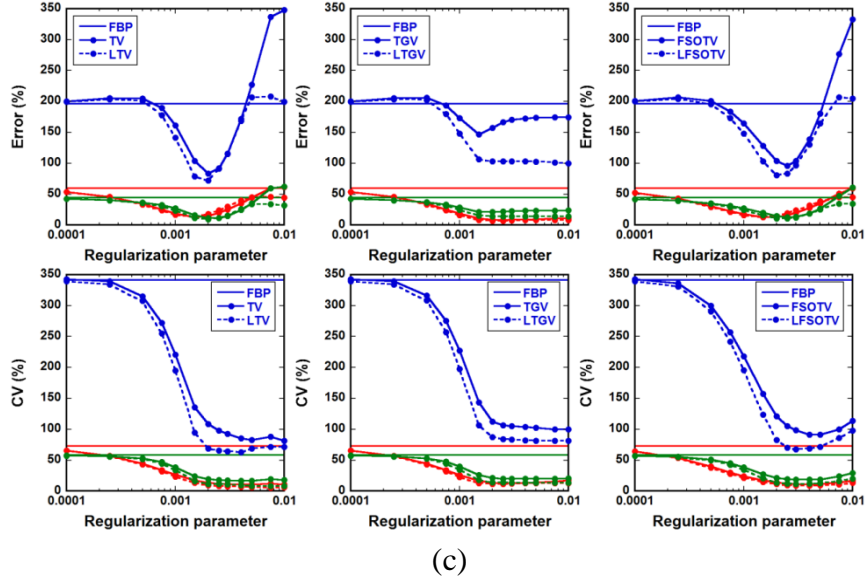

(c)

**Fig. 7.** *Error* (Equation (13)) (upper row) and *CV* (Equation (14)) (lower row) of estimated $K_{1a}$ (red), $K_{1p}$ (blue), and $k_2$ values (green) as functions of RP values for (a) non-cirrhotic CLD, (b) Child A cirrhosis, and (c) Child C cirrhosis (Table 1a). The left, middle, and right columns show the cases of TV (closed-circle-solid line) and LTV (closed-circle-dotted line), TGV (closed-circle-solid line) and LTGV (closed-circle-dotted line), and FSOTV (closed-circle-solid line) and LFSOTV (closed-circle-dotted line). Solid lines show cases for FBP. $I_0$ was assumed to be $5\times10^5$, and FBP input functions were used.

Figure 8a shows the $K_{1a}$, $K_{1p}$, and $k_2$ images (left to right columns) generated from the DCE-CT images reconstructed using FBP in the HCC model (Table 1b). The upper and lower rows show cases without and with noise, respectively. Figures 8b, 8c, and 8d show the $K_{1a}$, $K_{1p}$, and $k_2$ images, respectively, obtained using TV, LTV, TGV, LTGV, FSOTV, and LFSOTV (top to bottom rows) with RP values of 0.001, 0.002, 0.003, 0.004, and 0.005 (left to right columns). As shown in Figure 8a, the tumor was clearly recognized in all the HPP images without noise, whereas it was not recognized in the case with noise. As shown in Figure 8b, when using TGV and LTGV, an increase in $K_{1a}$ in the tumor was recognized at $\alpha_1 \geq 0.002$. When using the other regularizers, it was slightly recognized at $\alpha$ or $\alpha_1 = 0.002$, but not recognized at $\geq 0.003$. As shown in Figure 8c, a decrease in $K_{1p}$ in the tumor was recognized in all regularizers. However, when using regularizers except for TGV and LTGV, the contrast between the inside and outside of the tumor decreased with increasing RP values. As shown in Figure 8d, an increase in $k_2$ in the tumor was recognized in all regularizers with RP values $\geq 0.002$. When using the regularizers except for TGV and LTGV, the contrast between the inside and outside of the tumor increased with increasing RP values, whereas when using TGV and LTGV, it was almost constant at $\alpha_1 \geq 0.002$.

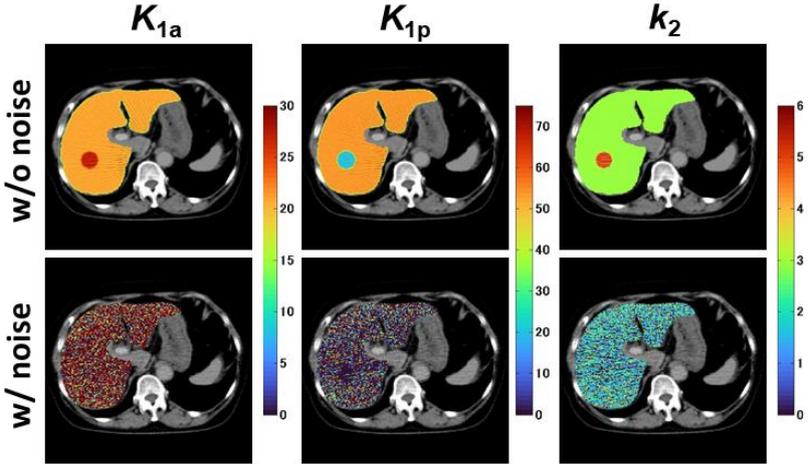

(a)

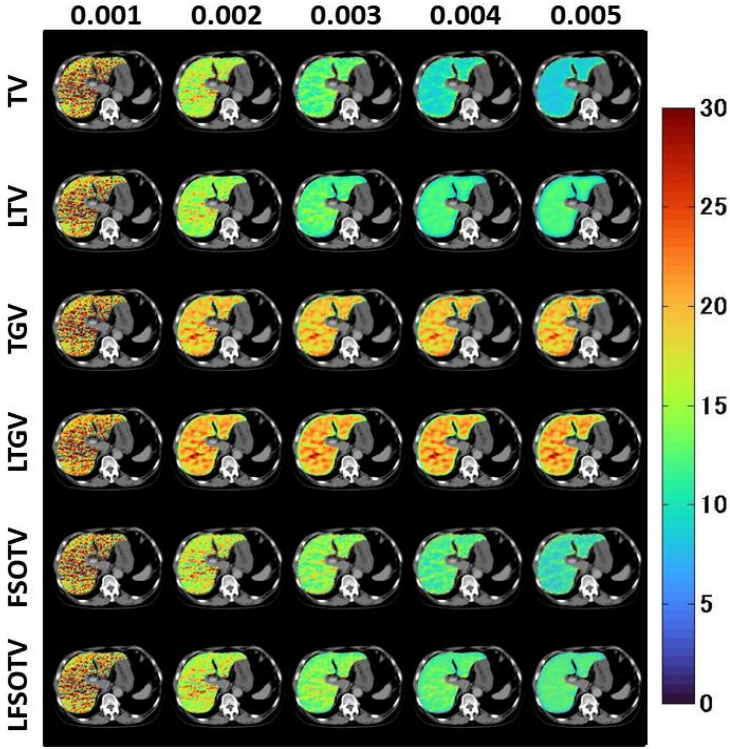

(b)

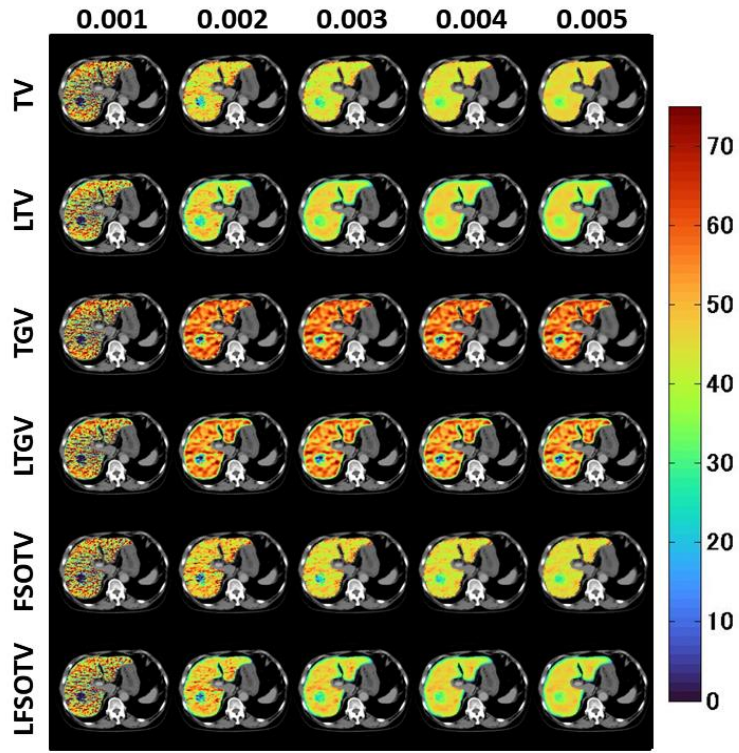

(c)

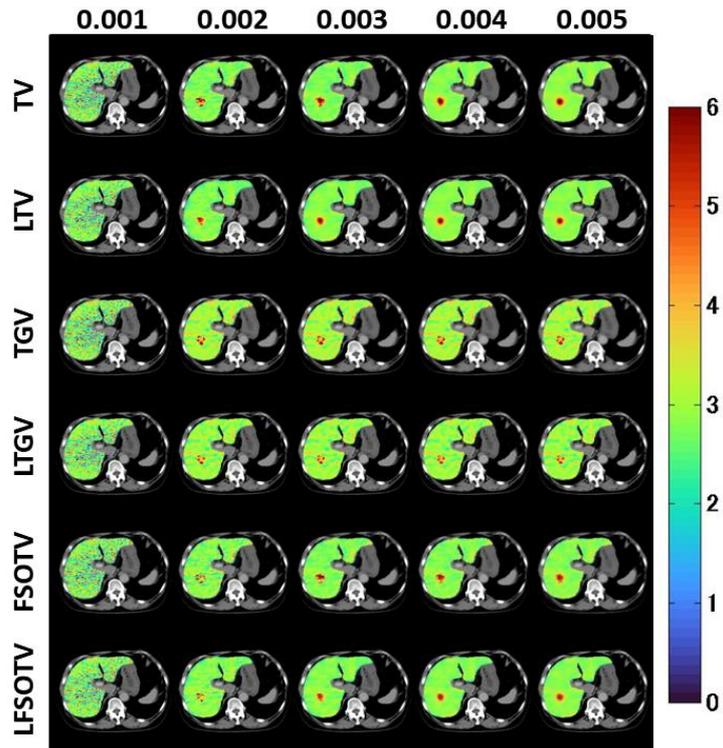

(d)

**Fig. 8.** (a) $K_{1a}$, $K_{1p}$, and $k_2$ images (left to right columns) generated from the DCE-CT images reconstructed using FBP in the HCC model (Table 1b). The upper and lower rows show cases without and with noise, respectively. (b) $K_{1a}$, (c) $K_{1p}$, and (d) $k_2$ images obtained by TV, LTV, TGV, LTGV, FSOTV, and LFSOTV (top to bottom rows) with RP values of 0.001, 0.002,

0.003, 0.004, and 0.005 (left to right columns). The $K_{1a}$, $K_{1p}$, and $k_2$ images are represented in mL/100mL/min, mL/100mL/min, and min$^{-1}$, respectively, and are superimposed on CT images. $I_0$ was assumed to be $5\times10^5$, and FBP input functions were used.

Figure 9 shows the tumor I/O ratios of $K_{1a}$ (red), $K_{1p}$ (blue), and $k_2$ (green) for the different regularizers as functions of RP values in the HCC model. The left, middle, and right panels show the cases of TV (closed circles) and LTV (dotted lines), TGV (closed circles) and LTGV (dotted lines), and FSOTV (closed circles) and LFSOTV (dotted lines). Tumor I/O ratios for FBP are shown as solid lines. The true tumor I/O ratios are shown in Table 1b. As shown in the left panel, when using TV, the tumor I/O ratio for $K_{1a}$ maximized at $\alpha = 0.0015$, following which it deceased. This decrease was slightly suppressed using LTV at $\alpha > 0.002$. That for $K_{1p}$ increased gradually with increasing $\alpha$ until 0.005, following which it plateaued and exceeded the true value. That for $k_2$ exhibited a peak at $\alpha = 0.0025$, following which it slightly decreased. When the LTV was used, this peak was slightly suppressed. When using TGV (middle panel), the tumor I/O ratio for $K_{1a}$ increased gradually with increasing $\alpha_1$ until 0.002, following which it plateaued. When using the LTGV, the plateau value increased slightly. That for $K_{1p}$ exhibited a peak at $\alpha_1 = 0.002$, following which it approached the true value. That for $k_2$ also exhibited a peak at $\alpha_1 = 0.002$, following which it plateaued. When using FSOTV and LFSOTV (right panel), the results were similar to those for TV and LTV (left panel).

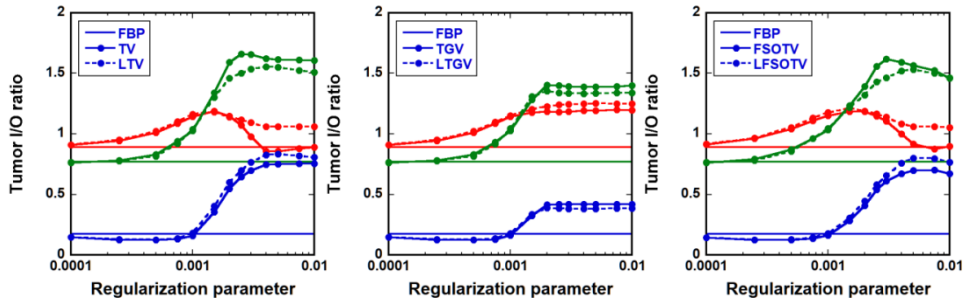

**Fig. 9.** Ratios of the mean $K_{1a}$ (red), $K_{1p}$ (blue), and $k_2$ values (green) inside and outside the tumor (tumor I/O ratios) obtained from the DCE-CT images reconstructed using different regularizers as functions of RP values in the HCC model (Table 1b). The left, middle, and right panels show the cases of TV (closed-circle-solid line) and LTV (closed-circle-dotted line), TGV (closed-circle-solid line) and LTGV (closed-circle-dotted line), and FSOTV (closed-circle-solid line) and LFSOTV (closed-circle-dotted line). Solid lines show the tumor I/O ratios for FBP. The true tumor I/O ratios are shown in Table 1b. $I_0$ was assumed to be $5\times10^5$, and FBP input functions were used.

The upper and lower rows of Figure 10 show the *Error* and *CV* of the estimated $K_{1a}$ (red), $K_{1p}$ (blue), and $k_2$ values (green) as functions of the RP values in the non-cirrhotic CLD (Table 1a) when $I_0$ was varied. The solid and dotted lines show the cases for $I_0 = 2.5\times10^5$ and $10^6$, respectively. The left, middle, and right panels show the TV, TGV, and FSOTV cases, respectively. To simplify the plots, the cases for LTV, LTGV, and LFSOTV are not shown. For $I_0 = 5\times10^5$, see Figure 7a. As shown in these figures, the RP values that minimized the *Error* and *CV* decreased with an increase in $I_0$.

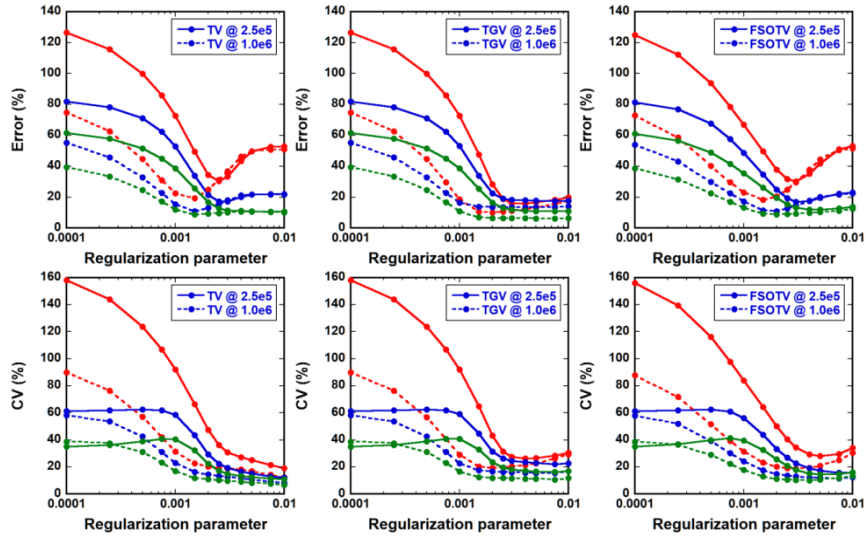

**Fig. 10.** *Error* (Equation (13)) (upper row) and *CV* (Equation (14)) (lower row) of estimated $K_{1a}$ (red), $K_{1p}$ (blue), and $k_2$ values (green) in non-cirrhotic CLD (Table 1a) as functions of RP values when $I_0$ was assumed to be $2.5\times10^5$ (closed-circle-solid line) and $1.0\times10^6$ (closed-circle-dotted line). The left, middle, and right columns show cases for TV, TGV, and FSOTV, respectively. In these cases, FBP input functions were used.

## 4 DISCUSSION

This study quantitatively investigated the performance of different simultaneous spatial and temporal regularizers with varying RP values when applied to low-dose DCE-CT liver perfusion studies by analyzing the generated HPP images. The results of this study will improve our understanding of regularizer characteristics in terms of HPP estimation accuracy and robustness. Additionally, this study will help select a suitable regularizer and/or RP value for low-dose DCE-CT liver perfusion studies.

The TGV expressed by Equation (6) is one of the regulariers using HOTV.[22] In the TGV, the contributions of the first- and second-order TVs were adjusted according to the image gradient ($\nabla_3 x$) via the auxiliary variable $v$ in Equation (6). When $v$ approaches zero, the contribution of first-order TV increases. In contrast, when it approaches $\nabla_3 x$, the contribution of the second-order TV becomes dominant. Therefore, TGV allows the preservation of sharp edges of images and piecewise smoothness, while allowing gradual image intensity changes. Thus, the TGV is expected to maintain the advantages of the TV while improving its drawbacks such as the occurrence of staircase artifacts.[26] For comparison, we also studied the performance of FSOTV. In FSOTV, the first- and second-order TVs are combined linearly (Equation (8)). This approach is aimed at preserving the advantages of the TV and reducing its drawbacks by adopting a fairly large and smaller weights for the first-order TV ($\alpha_1$) and second-order TV ($\alpha_2$), respectively. The major difference from TGV is that the ratio of these two TV contributions is fixed regardless of the image gradient in FSOTV. When $\alpha_2$ is zero, the FSOTV corresponds to the conventional TV. When $\alpha_2$ increases, the noise reduction due to the piecewise smoothness by the TV minimization decreases. In this study, the ratio of $\alpha_2$ to $\alpha_1$ was set to 0.1 by considering both the noise reduction and edge preservation effects.

When using the LRSD, DCE-CT images were decomposed into L and S components (Equation (4)). The L component of dynamic images can model the static features of a time-coherent (temporally correlated) background, including anatomical information. The S component can model the dynamic features.[27] Thus, the NN minimization of the L component is expected to help preserve static features owing to the reduction of background artifacts, whereas the TV minimization of the S component is expected to reduce noise in dynamic images.

In this study, the HPP values were estimated using a dual-input single-compartment model[2] and LLSQ method,[33] and their images were generated by applying this approach pixel by pixel. Furthermore, the arterial ($C_a(t)$) and portal venous input functions ($C_p(t)$) required for HPP quantification were extracted from the aorta and portal vein ROIs, respectively (Figure 1). Since these input functions may be affected by image reconstruction methods, we investigated their effects by calculating the AUCs of the input functions. Our results demonstrated that $C_p(t)$ was more affected by the regularizers than $C_a(t)$ (Figure 3). This was probably since $C_p(t)$ was more susceptible to the surrounding tissues via a partial volume effect than $C_a(t)$, and these effects differed depending on the regularizers. In addition, when using LRSD, the AUCs of $C_p(t)$ were smaller than those without LRSD (Figure 3). As previously described, the input functions were extracted from the CE images obtained by subtracting the pre-contrast (baseline) images from the DCE-CT images. As shown in Figure 2, the peak of $C_p(t)$ was much lower and broader than that of $C_a(t)$. Thus, the above finding may be because the separation of the L and S components obtained by LRSD in the portal vein ROI was inferior to that in the aorta ROI; thus, $C_p(t)$ was lower than that obtained without LRSD owing to excessive baseline subtraction.

As shown in Figures 4 and 5, when using individual input functions, the HPP values depended significantly on the image reconstruction methods. Of the HPPs, $K_{1p}$ was the most susceptible to differences in input functions. However, when the FBP input functions were used, the results were similar to those for the true input functions. In this study, we primarily used FBP input functions to exclude variations in the HPP estimation caused by the effect of regularizers on the input functions.

In most cases, the use of LRSD reduced the *Error* and *CV* of the estimated HPP values in the CLD model (Figure 7). In particular, LRSD in LTV or LFSOTV was useful for improving the performance of TV or FSOTV at large RP values in the case of Child C cirrhosis (Figure 7c). This finding can be attributed to the presevred static features of DCE-CT images, and their distortions caused by TV minimization (background artifacts) were suppressed using the LRSD at large RP values.

In CLDs, $K_{1p}$, $k_2$, and the sum of $K_{1a}$ and $K_{1p}$ decreased as the severity of CLD increased, whereas $K_{1a}$ tended to increase to compensate for the decrease in $K_{1p}$ levels (Table 1a). As shown in Figure 7, the *Error* and *CV* of the estimated HPP values changed significantly depending on the severity of the CLD. These results were consistent with those reported by Miyazaki et al.[39] In the Child C cirrhosis case with significantly reduced $K_{1p}$ and $k_2$, the *Error* and *CV* of the estimated $K_{1p}$ values were much larger than those for the other CLD cases (Figure 7c). When $K_{1p}$ and/or $k_2$ are significantly reduced as in the case of Child C cirrhosis, it may be effective to prolong the data acquisition duration of DCE-CT to improve HPP estimation accuracy, although this further burdens patients.

As shown in Figure 7, the *Error* and *CV* of $k_2$ were generally smaller than those of the other HPPs, suggesting robustness of $k_2$ against variations in $K_{1a}$ and $K_{1p}$. In CLDs, the MTT (reciprocal of $k_2$) increases as the intrahepatic vascular resistance increases due to the progression of hepatic fibrosis.[2] However, in HCCs, the intratumoral arteriovenous shunts increase compared to the surrounding non-tumor liver parenchyma, resulting in the decrease in MTT (increase in $k_2$).[3] As shown in Figure 8c, the increases in $k_2$ in the tumor and the contrast between inside and outside of the tumor were clearly observed. Thus, these results suggest that the $k_2$ value obtained in this study is useful for evaluating the progression of these pathophysiological states and may be helpful in differentiating between benign and malignant lesions.

When considering practical applications to low-dose DCE-CT perfusion studies, a regularizer that is insensitive and/or flexible to the choice of RP values is desirable as the ground truth used for evaluating the image quality and accuracy of perfusion parameter estimation is unknown, unlike in simulation studies. Thus, investigating the performance of the regularizers in detail is necessary when varying their RP values, as in this study. As shown in Figure 7, when using the regularizers other than TGV and LTGV in the CLD model, the *Errors* of the estimated HPPs were minimized at $\alpha$ or $\alpha_1 = 0.0015$–$0.002$, following which they increased. In contrast, when using TGV and LTGV, the *Errors* were minimized at $\alpha_1 = 0.0015$–$0.002$, following which they were approximately constant. This was confirmed by the visual inspection of the generated HPP images (Figure 6). When these regularizers were applied to the HCC model (Figures 8 and 9), the HPP images did not change significantly and the tumor I/O ratios were approximately constant at $\alpha_1 \geq 0.0015$–$0.002$. However, they changed depending on the RP values when using the other regularizers. These results suggest that the TGV and LTGV are more flexible in the choice of RP values as the ranges of their available RP values are larger than those of the other regulariers.

When $I_0$ was varied, the RP value minimizing the *Error* decreased with increasing $I_0$ and vice versa (Figures 7a and 10), although the dependencies of the *Error* on RP values were similar in all $I_0$ cases studied. Thus, these results suggest that when regularizers other than the TGV and LTGV are used, adjusting the RP value according to $I_0$ (the noise level) is necessary. However, achieving this in practical applications is difficult. In contrast, for the TGV and LTGV, if the RP value was set to that available at the lowest expected $I_0$, it could be applied even if $I_0$ was different from the expected value. Thus, TGV and LTGV appear to be superior to the other regularizers in terms of practicality.

However, striped artifacts were more remarkable in the HPP images obtained by the TGV and LTGV than in those obtained by the other regularizers (Figures 6 and 8). This is probably because the photon starvation artifacts remained owing to the edge preservation effect of the TGV and LTGV. Photon starvation artifacts are observed particularly in low-dose body CT and are location- and direction-dependent, that is, shift-variant.[42] Recently, Zeng et al.[42] have developed an effective nonlinear shift-variant method for reducing these artifacts in low-dose CT. The combination of the TGV and LTGV with this method may overcome the aforementioned drawbacks of these regularizers and enhance their usefulness. This study is currently ongoing.

## 5 CONCLUSIONS

We quantitatively investigated the performance of different simultaneous spatial and temporal regularizers in low-dose DCE-CT liver perfusion studies. Our results suggest that the LRSD and LTGV are useful for improving the accuracy of HPP estimation using low-dose DCE-CT and for enhancing its practicality. We expect this study to deepen our understanding of the regularizer performance and the selection of a suitable regularizer and/or RP value for low-dose DCE-CT liver perfusion studies.

## CONFLICT OF INTEREST STATEMENT

The authors declare no competing interest.

**REFERENCES**

1. Pandharipande PV, Krinsky GA, Rusinek H, Lee VS. Perfusion imaging of the liver: current challenges and future goals. *Radiology*. 2005;234(3):661–673.

2. Van Beers BE, Leconte I, Materne R, Smith AM, Jamart J, Horsmans Y. Hepatic perfusion parameters in chronic liver disease: dynamic CT measurements correlated with disease severity. *Am J Roentgenol*. 2001;176(3):667–673.

3. Kim SH, Kamaya A, Willmann JK. CT perfusion of the liver: principles and applications in oncology. *Radiology*. 2014;272(2): 322–344.

4. Leggett DA, Kelley BB, Bunce IH, Miles KA. Colorectal cancer: diagnostic potential of CT measurements of hepatic perfusion and implications for contrast enhancement protocols. Radiology 1997;205(3):716–720.

5. Miles KA. Measurement of tissue perfusion by dynamic computed tomography. *Br J Radiol*. 1991;64(761):409–412.

6. Hirata M, Sugawara Y, Fukutomi Y, Oomoto K, Murase K, Mochizuki T. Measurement of radiation dose in cerebral CT perfusion study. *Rad Med*. 2004;23(2):97–103.

7. Murase K, Nanjo T, Ii S, et al. Effect of x-ray tube current on the accuracy of cerebral perfusion parameters obtained by CT perfusion studies. *Phys Med Biol*. 2005;50(21):5019–5029.

8. Matasari R, Heryana N. Reduce noise in computed tomography image using adaptive Gaussian filter. *Int J Comput Tech*. 2019;6(1):17–20.

9. Mendrik AM, Vonken EJ, Van Ginneken B, et al. TIPS bilateral noise reduction in 4D CT perfusion scans produces high-quality cerebral blood flow maps. *Phys Med Biol*. 2011;56(13):3857–3872.

10. Murase K, Nanjo T, Sugawara Y, Hirata M, Mochizuki T. Usefulness of an anisotropic diffusion method in cerebral CT perfusion study using multi-detector row CT. *Open J Med Imaging*. 2015;5(3):106–116.

11. Zhang K, Zuo W, Chen Y, Meng D, Zhang L. Beyond a Gaussian denoiser: residual learning of deep CNN for image denoising. *IEEE Trans Image Process*. 2017;26(7):3142–3155.

12. Wolterink JM, Leiner T, Viergever MA, Išgum I. Generative adversarial networks for noise reduction in low-dose CT. *IEEE Trans Med Imaging*. 2017;36(12):2536–2545.

13. Chen H, Zhang Y, Zhang W, et al. Low-dose CT via convolutional neural network. *Biomed Opt Express*. 2017;8(2):679–694.

14. Wu D, Ren H, Li Q. Self-supervised dynamic CT perfusion image denoising with deep

neural networks. *IEEE Trans Radiat Plasma Med Sci*. 2021;5(3):350–361.

15. Lustig M, Donoho D, Pauly JM. Sparse MRI: the application of compressed sensing for rapid MR imaging *Magn Reson Med.* 2007;58(6):1182–1195.

16. Rudin LI, Osher S, Fatemi E. Nonlinear total variation based noise removal algorithms. *Physica D*. 1992;60:259–268.

17. Beck A, Tebouble M. Fast gradient-based algorithms for constrained total variation image denoising and deblurring problems. *IEEE Trans Image Process*. 2009;18(11):2419–2434.

18. Niu S, Zhang S, Huang J, et al. Low-dose cerebral perfusion computed tomography image restoration via low-rank and total variation regularizations. *Neurocomputing*. 2016;197:143–160.

19. Zeng D, Zhang X, Bian Z, et al. Cerebral perfusion computed tomography deconvolution via structure tensor total variation regularization. *Med Phys*. 2016;43(5):2091–2107.

20. Li S, Zeng D, Peng J, et al. An efficient iterative cerebral perfusion CT reconstruction via low-rank tensor decomposition with spatial–temporal total variation regularization. *IEEE Trans Med Imaging.* 2019;38(2):360–370.

21. Lysaker M, Lundervold A, Tai X-C. Noise removal using fourth-order partial differential equation with applications to medical magnetic resonance images in space and time. *IEEE Trans Image Process*. 2003;12(12):1579–1590.

22. Bredies K, Kunisch K, Pock T. Total generalized variation. *SIAM J Imaging Sci*. 2010;3(3):492–526.

23. Wang D, Smith DS, Yang X. Dynamic MR image reconstruction based on total generalized variation and low-rank decomposition. *Magn Reson Med*. 2020;83(6):2064–2076.

24. Wang C, Yin F-F, Kirkpatrick JP, Chang Z. Accelerated brain DCE-MRI using iterative reconstruction with total generalized variation penalty for quantitative pharmacokinetic analysis: a feasibility study. *Technol Cancer Res*. 2017;16(4):446–460.

25. Knoll F, Bredies K, Pock T, Stollberger R. Second order total generalized variation (TGV) for MRI. *Magn Reson Med*. 2011;65(2):480–491.

26. Huber R, Haberfehlner G, Holler M, Kothleitner G, Bredies K. Total generalized variation regularization for multi-modal electron tomography. *Nanoscale*. 2019;11(12): 5617–5632.

27. Otazo R, Candès E, Sodickson DK. Low-rank plus sparse matrix decomposition for

accelerated dynamic MRI with separation of background and dynamic components. *Magn Reson Med*. 2015;73(3):1125–1136.

28. Murase K, Nakamoto A, Tomiyama N. Simultaneous spatial and temporal regularization in low-dose dynamic contrast-enhanced CT cerebral perfusion studies. *J Appl Clin Med Phys*. 2023;24(6):e13983.

29. He J, Wang Y, Ma J. Radon inversion via deep learning. *IEEE Trans Med Imaging*. 2020;39(6):2076–2087.

30. Sidky EY, Jørgensen JH, Pan X. Convex optimization problem prototyping for image reconstruction in computed tomography with the Chambolle–Pock algorithm. *Phys Med Biol*. 2012; 57(10):3065–3091.

31. Xu J, Noo F. Convex optimization algorithms in medical image reconstruction – in the age of AI. *Phys Med Biol*. 2022; 67(7):10.1088/1361-6560/ac3842.

32. Murase K. Kashiwagi N, Tomiyama N. Quantitative evaluation of simultaneous spatial and temporal regularization in dynamic contrast-enhanced MRI of the liver using Gd-EOB-DTPA. *Magn Reson Imaging*. 2020;88:25-37.

33. Murase K, Miyazaki S, Yang X. An efficient method for calculating kinetic parameters in a dual-input single-compartment model. *Br J Radiol*. 2007;80(953):371–375.

34. Brix G, Griebel J, Kiessling F, Wenz F. Tracer kinetic modelling of tumour angiogenesis based on dynamic contrast-enhanced CT and MRI measurements. *Eur J Nucl Med Mol Imaging*. 2010;37:S30–S51.

35. Miyazaki S, Murase K, Yoshikawa T, et al. A quantitative method for estimating hepatic blood flow using a dual-input single-compartment model. *Br J Radiol*. 2008;81(970):790–800.

36. Iwamoto S. X-ray attenuation coefficient in diagnostic imaging. *Jpn J Imaging Inf Sci Med*. 2015;32(3):54–62.

37. Kak AC, Slaney M. Principles of Computerized Tomographic Imaging (Piscataway NJ: IEEE Press) Reprint: SIAM Classics in Applied Mathematics, 2001;49–112.

38. Siddon RI. Fast calculation of the exact radiological path for a three-dimensional CT array. *Med Phys*. 1985;12(2):252–255.

39. Miyazaki S, Yamazaki Y, Murase K. Error analysis of the quantification of hepatic perfusion using a dual-input single-compartment model. *Phys Med Biol*. 2008;53(21):5927–5946.

40. Chen Y-W, Pan H-B, Tseng H-H, Hung Y-T, Huang J-S, Chou C-P. Assessment of blood flow in hepatocellular carcinoma: correlations of computed tomography perfusion imaging and circulating angiogenic factors. *Int J Mol Sci*. 2013;14:17536–17552;

41. Sahani D, Holalkere N-S, Mueller PR, Zhu AX. Advanced hepatocellular carcinoma: CT perfusion of liver and tumor tissue—initial experience. *Radiology*. 2007;243(3):736–743.

42. Zeng GL. Photon starvation artifact reduction by shift-variant processing. *IEEE Access*. 2022;10:13633–13649.